\documentclass[conference]{IEEEtran}
\ifCLASSINFOpdf
\else
\fi
\usepackage[pdftex]{graphicx}
\usepackage{float}
\usepackage{graphicx,epstopdf}
\usepackage{amsmath,amssymb,amsfonts,subfigure}
\usepackage{comment}
\usepackage{adjustbox}
\usepackage{lipsum}
\usepackage{cite}

\begin{document}
%
\title{Evolving Models for Meso-Scale Structures}

\author{\IEEEauthorblockN{Akrati Saxena}
\IEEEauthorblockA{Computer Science and Engineering\\
Indian Institute of Technology
Ropar, India\\
Email: akrati.saxena@iitrpr.ac.in}
\and
\IEEEauthorblockN{S. R. S. Iyengar}
\IEEEauthorblockA{Computer Science and Engineering\\
Indian Institute of Technology
Ropar, India\\
Email: sudarshan@iitrpr.ac.in}}

\maketitle

\begin{abstract}
Real world complex networks are scale free and possess meso-scale properties like core-periphery and community structure. We study evolution of the core over time in real world networks. This paper proposes evolving models for both unweighted and weighted scale free networks having local and global core-periphery as well as community structure. Network evolves using topological growth, self growth, and weight distribution function. To validate the correctness of proposed models, we use K-shell and S-shell decomposition methods. Simulation results show that the generated unweighted networks follow power law degree distribution with droop head and heavy tail. Similarly, generated weighted networks follow degree, strength, and edge-weight power law distributions. We further study other properties of complex networks, such as clustering coefficient, nearest neighbor degree, and strength degree correlation.
\end{abstract}

\IEEEpeerreviewmaketitle

\section{Introduction}

Complex systems have been a part of many research areas, such as biological networks, chemical networks, technological networks, business networks. A complex network can be represented as a graph $G(V,E)$, where the vertex set $V$ represents the set of objects, and the edge set $E$ represents the relationships between these objects. Each edge in $E$ is represented by an ordered/unordered pair $(x,y)$ for directed/undirected graphs, where $x,y \in V$. Such networks can also be represented by an adjacency matrix $A$, where $a_{ij}^{th}$ entry of A is $1$, if node $i$ is connected to node $j$ otherwise $0$. Degree of a node denotes the number of edges connected to a node and can be defined as $d_i=\sum_{j \epsilon \Gamma (i)} a_{ij}$, where $\Gamma(i)$ is the set of neighbors of node $i$.

The binary links do not convey complete information about the system. In 1973, Granovetter emphasized the idea of inequality of the edges in a network \cite{granovetter1973strength}. He broadly categorized edges as weak ties and strong ties. In real world complex networks, each link carries a unique weight \cite{latora2001efficient}. Edge weights portray a better understanding of the network. For example, In friendship networks, edge weights can be used to denote the intimacy of the relationship or frequency of  communication \cite{onnela2007analysis}. Edge weights can also represent the number of publications in a collaboration network or traffic flow in an airport network \cite{de2005structure}.

In real world networks, edges can also carry negative weights to represent varying degrees of friendship-enmity \cite{brzozowski2008friends} or trust-distrust \cite{guha2004propagation}. Weighted networks are characterized by a function $f(E) \rightarrow R$. The function $f$ maps each edge to a real number, representing weight of that edge. Such networks can be stored as a weighted adjacency matrix, where $w_{ij}$ denotes the weight of an edge connecting nodes $i$ and $j$. As a generalization of degree, strength of a node can be defined as, $s_i=\sum_{j \epsilon \Gamma (i)} w_{ij}$.

All real world complex networks follow some macro level properties, such as six degrees of separation (small world effect) \cite{milgram1967small}, scale free degree distribution \cite{barabasi1999emergence}, preferential attachment \cite{newman2001clustering}, high clustering coefficient \cite{newman2003structure}.

Topological structure of complex networks also have meso scale properties, which include community structure and core-periphery structure. In the community structure, each node has some links in its own community called \textit{intra community links} and some links to other communities called \textit{inter community links}. Probability of having an intra community link is higher than the inter community link. In  weighted networks, intra community links tend to have higher weights than inter community links.

Hierarchical organization of the network gives birth to the core-periphery structure that coexists with the community structure. Concept of core-periphery structure was first proposed by Borgatti and Everettee \cite{borgatti2000models}. Core is a densely connected nucleus of the network that is surrounded by sparsely connected periphery nodes. Every community has a local core and all these local cores lead to the emergence of global core in the complex network. Seidman \cite{seidman1983network} proposed K-shell decomposition method to identify core nodes in an unweighted network. Eidsaa and Almaas \cite{eidsaa2013s} extended K-shell decomposition method and proposed S-shell decomposition method for weighted networks. Saxena et al. show the importance of core nodes in information diffusion on the networks having meso scale structures \cite{saxena2015understanding}.

In the present work, we study evolution of the core in real world networks i.e. how core nodes acquire a well known position in the network. We also propose evolving models for both unweighted and weighted networks having community and core-periphery structure. The rest of the paper is organized as follows. In section 2, we review the literature. This is followed by preliminaries and datasets defined in section 3 and 4 respectively. Section 5 explains evolution process of the core in real world networks. In section 6, we propose evolving models along with simulation results to validate different properties of real world networks. Section 7 is dedicated to the conclusion and scope for future work.

\section{Related Work}

Various modeling frameworks have been designed to understand the topological structure of dynamically growing complex networks. This line of research was ignited by Erdos-Renyi, when they proposed a very basic first generative model for the structure of real world complex networks \cite{erdds1959random}. Their model portrayed the real world network as a random network, where each pair of nodes are connected with a fixed probability. This model was inadequate in explaining all the properties of a real world network. In 1998, Watts and strogatz \cite{watts1998collective} designed a model to better explain the small world phenomenon of real world social networks. They considered that friendships are not random, rather they are based on the conscious decision of random encounters. They emphasized that the clustering coefficient of a real world network is large, with a small diameter.

In 1999, Barabasi-Albert studied the dynamics of growing networks, where new nodes arrive and make connections with existing nodes. They empirically observed that most of the real world networks follow a power law degree distribution. They proposed a preferential attachment model based on \textit{rich-get-richer} phenomenon, which results in a scale free network having few hubs \cite{barabasi1999emergence}. This work added a new dimension to the study of complex networks.

After this pioneering work, many other evolving models have been proposed based on different node attributes and evolving environment. We can classify those models into few main categories. P L Krapivsky et al. \cite{krapivsky2000connectivity} proposed a non-linear preferential attachment model, where the probability to get new connections for a node of degree $k$ is $P(k) = k^y$, where, $y$ is a constant and its different values yield different type of scale free networks. Dorogovtsev and Mendes \cite{dorogovtsev2000effect} studied the role of accelerating growth on network structure. They also \cite{dorogovtsev2000evolution} proposed a model, where probability of a connection depends on both degree as well as age of the nodes.

In 2001, Bianconi and Barabasi \cite{bianconi2001competition} proposed a \textit{fitter-get-richer} modeling phenomenon, where probability of connection is directly proportional to the fitness value of a node. There are some more models based on different properties of nodes, such as intrinsic strength \cite{geng2007weighted}, competitiveness \cite{guo2015universality}, initial attractiveness \cite{dorogovtsev2000structure}, or capacity constraints \cite{wu2013weighted}. Jost and Joy \cite{jost2002evolving} proposed an evolving model for social networks, where each person prefers to be friend with the friends of its friends. Friendship between two persons also depends on their mutual affinity, intimacy, attachment and understanding between them. WX Wang et al. \cite{wang2006mutual} proposed a model based on the mutual attraction. \cite{li2003local, sun2007unweighted} have proposed local world models based on the concept of local world connectivity. \cite{gastner2006spatial, barrat2005effects} studied geographical constraint model, where a network is embedded in a two-dimensional geographic plain. \cite{li2004comprehensive, wang2005mutual} proposed a self growing model, where existing nodes also make new connections in the future.

In traffic networks, when a new node connects to an existing node, it increases traffic on that node. This increased traffic is distributed across all neighbors of the node. Such a distribution is called local rearrangement of the weights. In 2004, A Barrat et al. \cite{barrat2004modeling} studied this approach and proposed a weighted evolving model, that is also called BBV (Barrat, Barthelemy, and Vespignani) model. \cite{dai2013weighted, zheng2008weighted} proposed traffic driven weighted models, where edge weights can be calculated using traffic flow or betweenness centrality measure.

Researchers have also explained the evolution of meso scale properties, such as, community structure, hierarchical organization, multilayer networks. \cite{pollner2006preferential, li2005evolving, xie2007new} proposed evolving models for community structure in  unweighted networks. Newman \cite{newman2004analysis} studied the community structure and modified modularity function to quantify the quality of community structure for weighted networks. \cite{li2006modelling, kumpula2007emergence, zhou2008weighted} proposed generative models for weighted community structured complex networks.

Unlike community structure, core-periphery structure has not been deeply explored. 
In this work, we analyze the evolving phenomenon of core-periphery structure in real world networks. We also propose evolving models for complex networks having community structure along with local and global core-periphery structure. As per the best of our knowledge, this is the first work of its kind. This model will help the community of network science researchers to understand the structure and dynamics of networks at the level of information diffusion, epidemic spread, and spread of influence. 

\section{Preliminaries}

\subsection{Preferential Attachment}

Preferential attachment model is an evolutionary model proposed by Barabasi and Albert for the formation of unweighted scale free networks \cite{barabasi1999emergence}. In this model, probability $\prod (k_x)$ of a new node $i$ connecting to an existing node $x$ depends on the degree $k_x$ of node $x$.

\begin{center}
$\prod (k_x) = \frac{k_x}{\sum_{y}k_y}$
\end{center}
So, the nodes having higher degrees acquire more links over time.

Analogously, in weighted networks, nodes follow preferential attachment in accordance with the strength $s_x$ of nodes. It is defined as,

\begin{center}
$\prod (s_x) = \frac{s_x}{\sum_{y}s_y}$
\end{center}
As we can see, both type of networks follow a \textit{rich-get-richer} phenomenon, thereby skewing the distribution towards lower degree.

\subsection{Power Law Distribution}

Preferential attachment model gives rise to power law degree distribution, where probability of a node having degree $k$, $P(k)$ is proportional to $ k^{-\gamma}$. As the network grows, only a few nodes, called hubs, manage to get a very large number of links.

Weighted networks also follow three power law distributions: 1. Power law degree distribution, 2. Power law strength distribution, and 3. Power law edge-weight distribution.


\subsection{K-shell Decomposition Algorithm}

K-shell algorithm \cite{seidman1983network} is a well known method in social network analysis to find the tightly knit group of influential core nodes in a given network. This method divides the entire network into shells and assigns a shell index to each node. The innermost shell has the highest shell index $k_{max}$ and is called nucleus of the network.

This algorithm works by recursively pruning the nodes from lower degree to higher degree. First, we recursively remove all nodes of degree 1, until there is no node of 1 degree. All these nodes are assigned shell index $k_s=1$. In similar fashion, nodes of degree 2,3,4... are pruned step by step. When we remove nodes of degree $k$, if there appears any node of degree less than $k$, it will also be removed in the same step. All these nodes are assigned shell index $k$. Here, higher shell index represents higher coreness.

Eidsaa and Almaas \cite{eidsaa2013s} modified K-shell decomposition algorithm for weighted networks. They consider strength instead of degree for dividing nodes into various shells. It is called S-shell or strength decomposition algorithm.

\section{Datasets}

We use following undirected network datasets to understand the evolution and structure of the core:
\begin{enumerate}
\item ArXiv hep-ph Collaboration network: This is the co-authorship network of ArXiv's High Energy Physics Phenomenology (hep-ph) publications \cite{leskovec2005graphs}. We create two networks for different time periods to understand evolution of the core nodes in the network:
\begin{enumerate}
\item ArXiv hep-ph(1) Collaboration network: It contains 16,959 nodes and 1,194,440 edges. It includes all publications from timestamp 0 to 944780400.
\item ArXiv hep-ph(2) Collaboration network: It contains 28,093 nodes and 3,148,447 edges. It includes all publications from timestamp 0 to 1015887601.
\end{enumerate}
\item Facebook: Facebook is the most popular online social networking site today. This dataset is the induced subgraph of Facebook \cite{viswanath2009evolution}, where users are represented by nodes and friendships are represented by edges.
\begin{enumerate}
\item Facebook(1) dataset: It contains 61,185 nodes and 761,777 edges. It includes all friendships existing from timestamp 0 (starting of the network) to 1227000000.
\item Facebook(2) dataset: It contains 63,731 nodes and 817,035 edges. It includes all friendships existing from timestamp 0 to 1230998204.
\end{enumerate}
\item DBLP Collaboration network: This is a weighted coauthorship network extracted from DBLP computer science bibliography, where edge weight denotes the number of common publications\footnote{Dataset is downloaded from http://projects.csail.mit.edu/dnd/DBLP/}. 
\begin{enumerate}
\item DBLP(1) Collaboration network: This network contains 1,345,696 nodes and 5,560,540 edges. It includes all publications from starting to 2013.
\item DBLP(2) Collaboration network: This network contains 1,406,703 nodes and 5,927,742 edges. It includes all publications upto April 2015.
\end{enumerate}
\item ArXiv hep-th Collaboration network: This is the co-authorship network of ArXiv's High Energy Physics Theory (hep-th) \cite{leskovec2005graphs}. The network contains 12,309 nodes and 154,769 edges.
\item Google plus: Google plus is an online social networking facility provided by Google Services. This network contains 107,614 nodes and 12,238,285 edges \cite{leskovec2012learning}.
\item Twitter: This is an induced subgraph of Twitter and it contains 81,306 nodes and 1,342,310 edges \cite{leskovec2012learning}.
\item Enron email network: This is an email communication network of Enron organization between 1999 and 2003 \cite{klimt2004introducing}. The dataset has 87,273 nodes and 299,220 edges.
\item Actor Collaboration Network: In this network, nodes are the actors and there is an edge between two actors if both of them have appeared in the same movie \cite{barabasi1999emergence}. It has 382,219 nodes and 15,038,083 edges.
\end{enumerate}

\section{Evolution of Core Structure}

Real world networks have self regulatory evolving phenomenon that gives rise to a core-periphery structure. Core is a very dense nucleus of the network that is highly connected with periphery nodes. In social networks, core nodes are the group of highly elite people of the society. Similarly in co-authorship network, core nodes are the pioneer researchers of the area.

In the present work, we use K-shell and S-shell decomposition methods to identify core for unweighted and weighted networks respectively. We analyze distribution of the nodes in different shells. Table 1 shows that the core constitutes less than 2 percent of the whole network. Gupta et al. \cite{gupta2015pseudo} observed that few innermost shells close to nucleus have the same influential power as core and have very high density. In practical use, a very small percentage of nodes from innermost layers can also be considered as core nodes.

\begin{table}[]
\centering
\caption{Fraction of core nodes in real world networks}
\begin{adjustbox}{width=.5\textwidth}
\label{my-label1}
\begin{tabular}{|l|l|l|l|l|}
\hline
Dataset & \multicolumn{1}{|p{1.7cm}|}{\centering Number of \\Total Nodes}  & \multicolumn{1}{|p{1.7cm}|}{\centering Number of \\ Core Nodes} & \multicolumn{1}{|p{1.7cm}|}{\centering Fraction of \\ Core Nodes (in $\%$)} & \multicolumn{1}{|p{1.7cm}|}{\centering Density of \\ Core}\\ \hline
ArXiv hep-ph(2) & 28093 & 411 & 1.46 & 1.0 \\ \hline
Facebook(2) & 63731 & 699 & 1.1 & 0.126 \\ \hline
DBLP(2) & 1406703 & 119 & 0.008  & 1.0 \\ \hline
ArXiv hep-th & 12309 & 83 & 0.67 & 1.0 \\ \hline
Google Plus & 107614 & 1865 & 1.73 & 0.5535 \\ \hline
Twitter & 81306  & 130 & 0.16 & 0.907 \\ \hline
Enron & 87273 & 102 & 0.12 & 0.652 \\ \hline
Actor Collaboration & 382219 & 1178 & 0.31 & 0.47 \\ \hline
\end{tabular}
\end{adjustbox}
\end{table}

In this study, we examine evolution of the core over time by using ArXiv hep-ph and Facebook timestamped unweighted networks. We create two networks for different time spans, where second network includes more number of nodes and edges as explained in dataset section. The shell structure of all these networks is determined using K-shell decomposition algorithm. We find that in second network of both the datasets, some new nodes are added to the core. We specifically analyze properties of these nodes with respect to other nodes, which were also having same degree in the first network but could not shift to the core. Study shows that the networks evolve using preferential attachment. Each node competes to get more and more connections to become part of the center of the network. We have observed that when a node competes and acquires some high number of connections with other core nodes, this node is also shifted to core. Along with that, periphery nodes are also more attracted to make connections with this node. This sudden hike in the number of connections with other core nodes helps this node to be a part of the core. A node is converted into a core node with very less probability. Similar results are observed by creating networks for different time spans by increasing the time period with a small time gap.

\begin{figure}[]
\centering
\includegraphics[width=8cm]{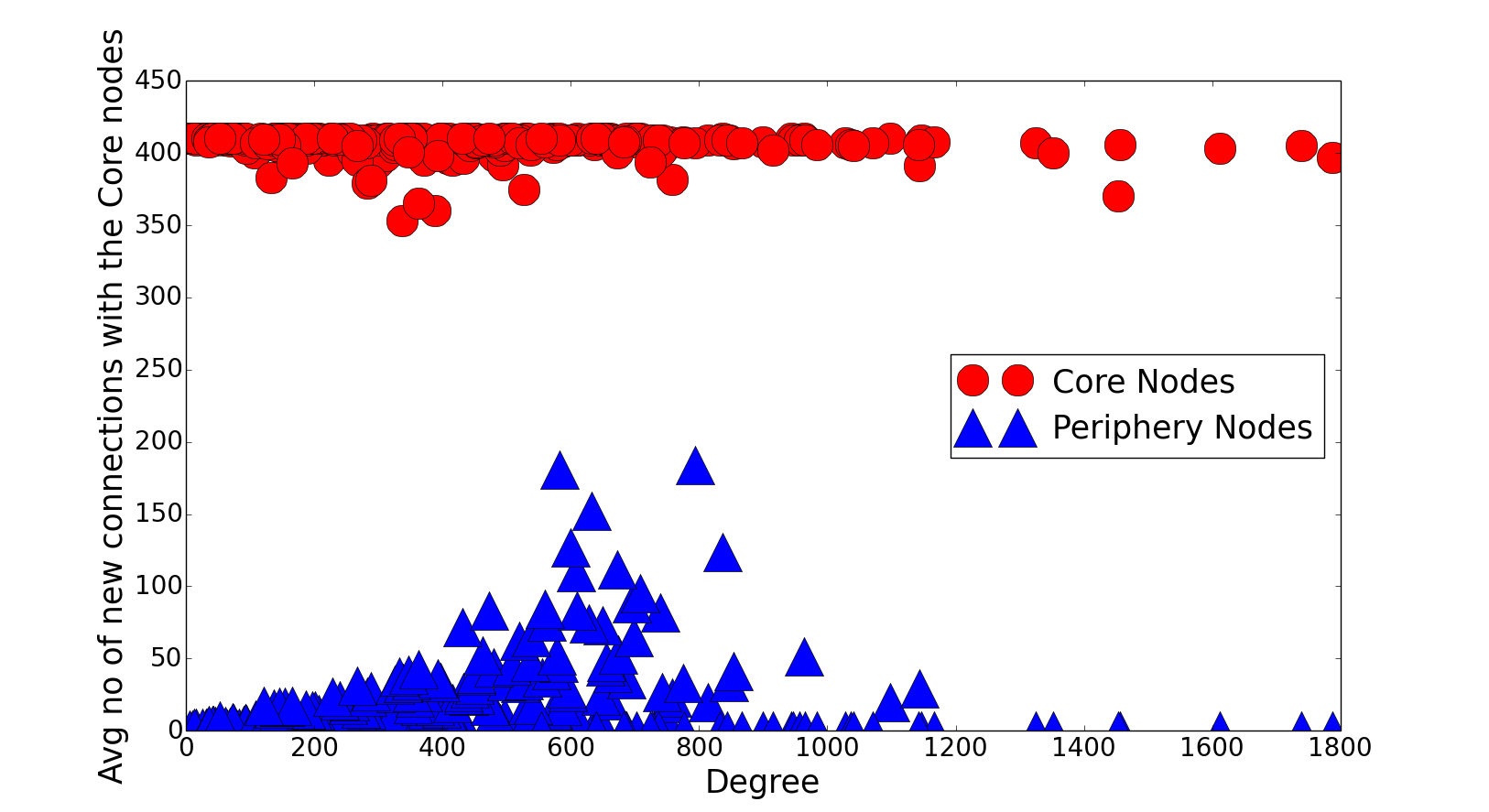}
\caption{Average increase in number of connections with core nodes for ArXiv hep-ph dataset, where circles represent the nodes shifted to the core and triangles represent the nodes which remain in periphery}
\end{figure}

\begin{figure}[]
\centering
\includegraphics[width=8cm]{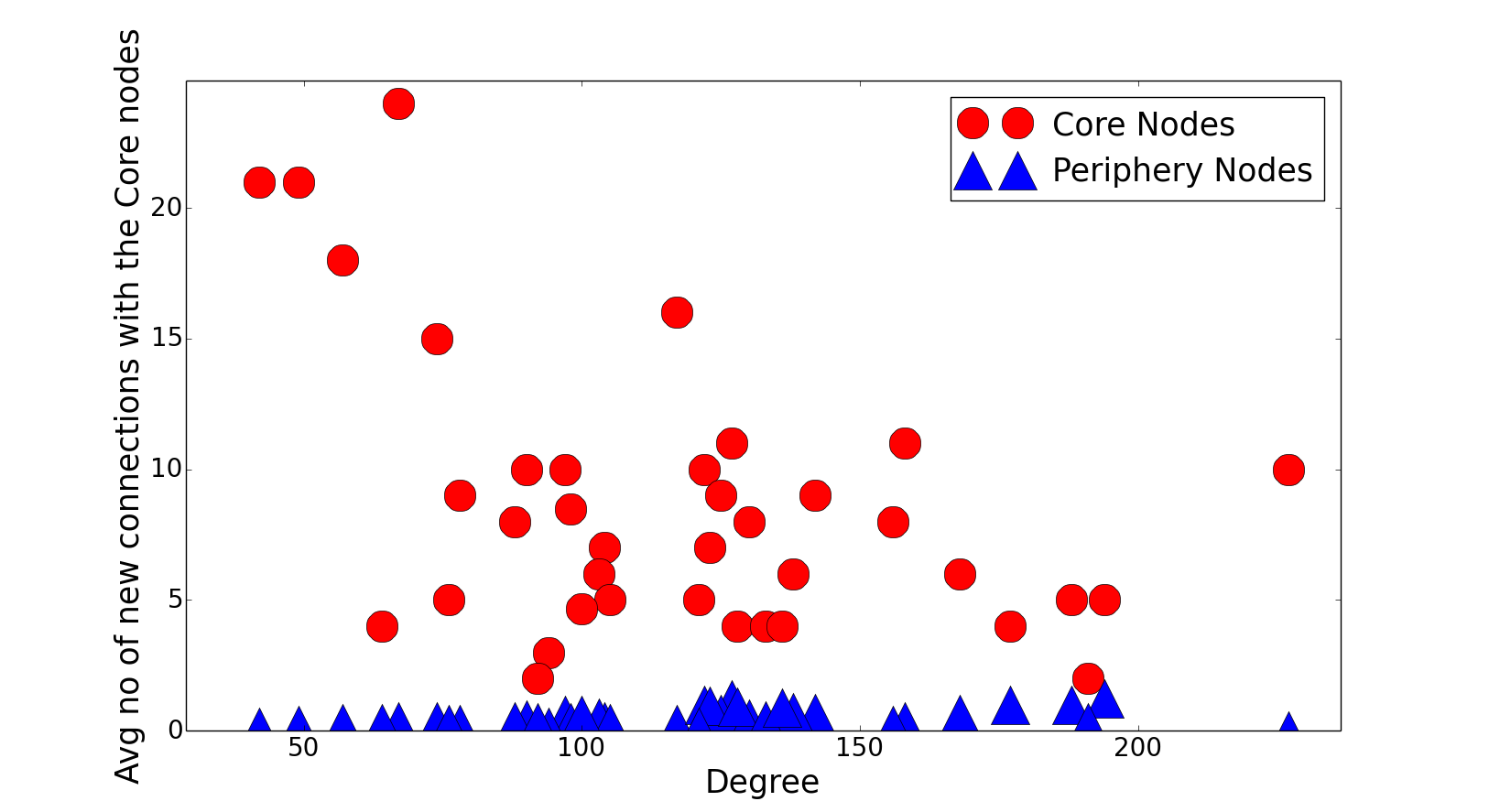}
\caption{Average increase in number of connections with core nodes for facebook dataset, where circles represent the nodes shifted to the core and triangles represent the nodes which remain in periphery}
\end{figure}

This phenomenon is explained in Fig. 1 for ArXiv hep-ph datasets. Red color circles show the average increase in number of connections with other core nodes, when a node is shifted to core from periphery. Blue color triangles show the average increase in core connections for nodes having same degree as the shifted node but they remain in the periphery. It shows that when a node gets connected with many core nodes, it is converted into a core node. This sudden growth in the connections gives birth to a small and dense core structure present in the network. Similar results are shown for the Facebook dataset in Fig. 2.

\begin{figure}[]
\centering
\includegraphics[width=8cm]{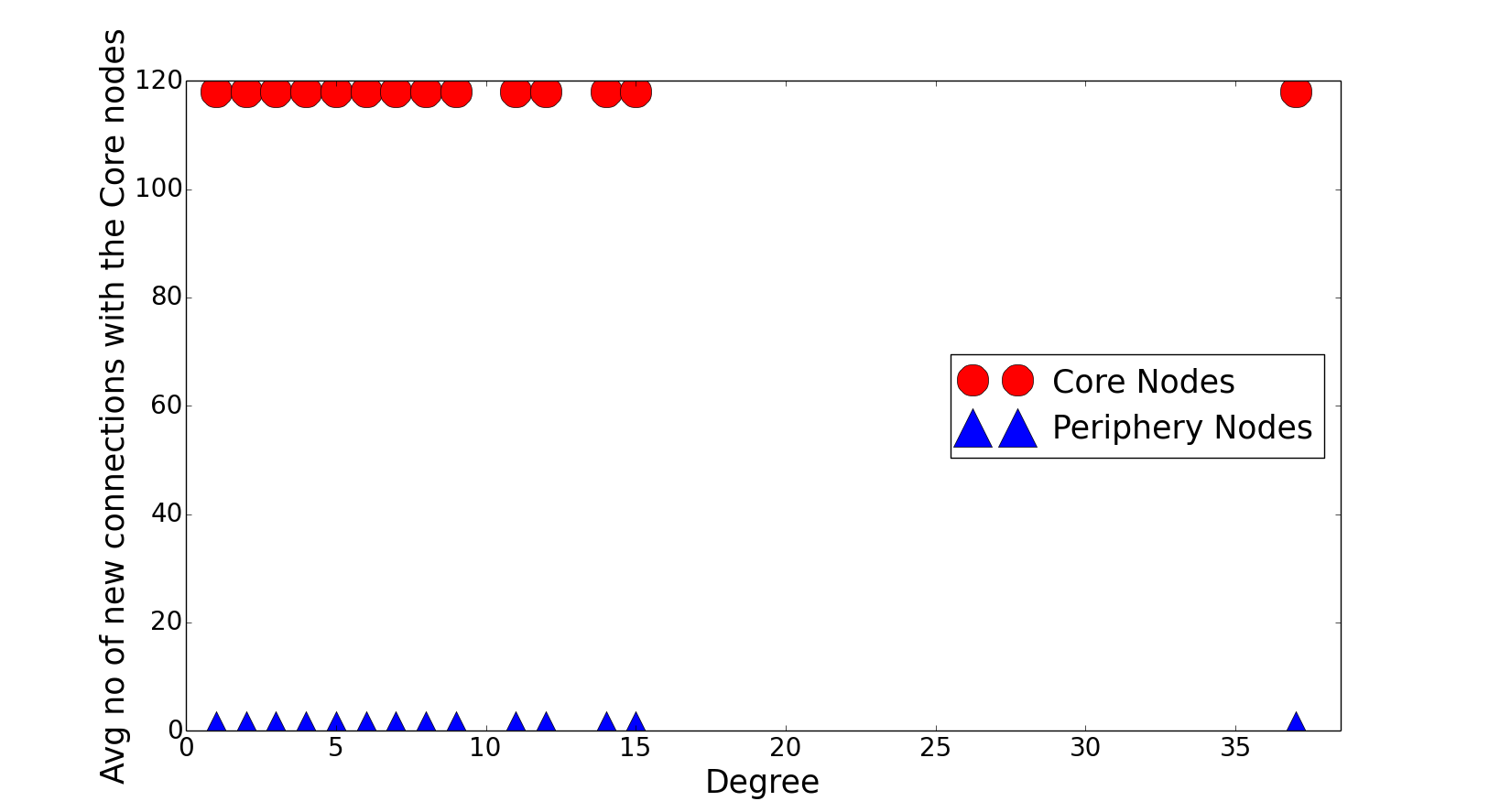}
\caption{Average increase in number of connections with core nodes for DBLP dataset, where circles represent the nodes shifted to the core and triangles represent the nodes which remain in periphery}
\end{figure}

For weighted networks, we use DBLP dataset and perform S-shell decomposition method. We create networks for different time spans with the time gap of one year and observe similar results, as shown in Fig. 3.

We also study average increase in total number of connections for both cases, when a node is shifted to the core or when it remains in the periphery. Results show that the increase in total number of connections is not a good indicator of the coreness of a node. It proves that total increase in core connections plays an important role and modulates growth of the core. Fig. 4 shows these results for ArXiv hep-ph datasets. Similar results are observed for Facebook and DBLP datasets.

\begin{figure}[]
\centering
\includegraphics[width=8cm]{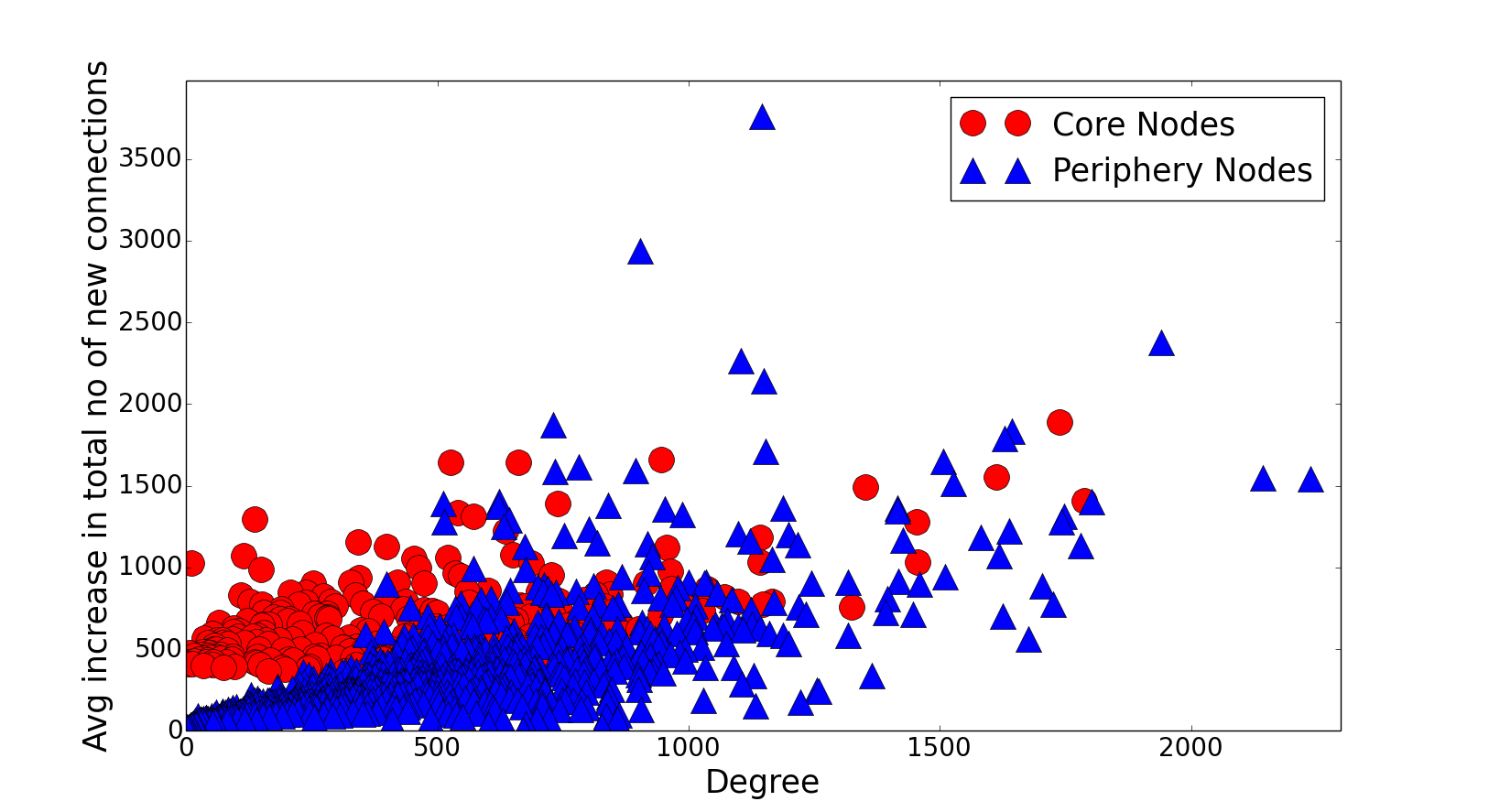}
\caption{Average increase in total number of connections for ArXiv hep-ph dataset, where circles represent the nodes shifted to the core and triangles represent the nodes which remain in periphery}
\end{figure}                                                                                                                          
\section{The Proposed Model}
We propose evolving models based on the properties of community and core-periphery structure for both unweighted and weighted scale free networks. A community $i$ is denoted as $C_i$ and global core is denoted as $Core$. If a node $x$ $\epsilon$ $Core$, its type is set as \textbf{core}, otherwise \textbf{periphery}. Models are based on topological growth and self growth process. In topological growth, network expands as new nodes keep joining. In real world networks, all existing nodes also keep making new connections among them. This self growing phenomenon is captured by the self growth process.

Models evolve using two types of links:
\begin{enumerate}
\item Preferential Attachment link: In preferential attachment, probability of a node getting a new connection is directly proportional to its degree. In our models we use following preferential attachment rules:
\begin{enumerate}
\item Intra-Community Preferential Attachment: When a new node $x$ joins a community $C_i$, probability of this new node making connection with a node $y$ in its own community is directly proportional to intra-degree of $y$ with respect to other nodes in the community. It is defined as,
\begin{center}
$\prod(y)=\frac{k^1_y}{\sum_{z \epsilon C_i}k^1_z}$
\end{center}
where, $k^1_{y}$ is the intra-degree of node $y$ in community $C_i$. It denotes the number of neighbors belonging to its own community.
\item Inter-Community Preferential Attachment: Probability of a new node $x$ $(x \epsilon C_i)$ making inter community connection with a node $y$ $(y \epsilon C_j)$ is proportional to inter-degree $k^2_y$ of node $y$. It is defined as,
\begin{center}
$\prod(y)=\frac{k^2_y}{\sum_{z \notin C_i}k^2_z}$
\end{center}
\end{enumerate}
\item Triad Formation Links: When an existing node makes new connections with the friends of its friends, these links are called Triad Formation (TF) links. To make a TF link, node $x$ selects one of its neighbor $y$ preferentially and then one neighbor $z$ of $y$ is chosen preferentially that is already not connected with $x$. TF links are important and moderate the self growth of the network.
\end{enumerate}

In weighted networks, strength is used in place of degree for all preferential attachment functions. We also use weight distribution function proposed by BBV model to balance the load. In section 6.1, we present our model for unweighted networks and discuss its simulation results in next section. Model is modified to generate weighted networks and is discussed in section 6.3 and 6.4. In section 6.5, efficiency of the proposed models is calculated for the generated networks of different sizes.

\subsection{Model A}

Network starts with a seed graph having $c$ number of communities. Each community has clique of size $n_0$. In the seed graph, each node is connected to each inter community node with a very high probability $p$. All nodes of the seed graph are set as core nodes. Every newly coming node makes $f$ fraction of connections in its own community and $(1-f)$ fraction of connections with other communities. $f$ controls the density of intra and inter community links. At every time stamp, the steps followed are:
\begin{enumerate}
\item A new periphery node $x$ is added to a uniformly at randomly selected community. This new node makes $m$ connections in total. $m \cdot f$ intra community links are connected using intra-community preferential attachment and $m \cdot (1-f)$ inter community links are connected using inter-community preferential attachment.
\item With probability $q$, a periphery node is selected using degree preferential rule and it is converted into a core node. This node is connected to each core node with probability $p$. This is an important step to capture the evolution process of core nodes. 
\item Each node makes a triad link with probability $r$ using triad formation rule defined in the last section.
\end{enumerate}

In this model, step 1 controls the topological growth, and step 2 and 3 control the self growth of the network. Evolution of the global core is controlled by step 2. Probability $p$ depends on the density of the core and its value is very high. Probability $q$ controls size of the core that includes the nucleus and few innermost tightly knit and highly dense layers of the network. Probability $r$ controls the rate of the self growth and its value is very small, since existing nodes make less number of connections with other nodes as the network size grows.

\subsection{Simulation Results}

We perform network simulation by varying different parameters of the model. All results are shown for the network having 20,000 nodes and 826,759 edges. Seed graph contains $c=6$ communities and each community is started with the clique of $n_0=3$ nodes. All inter community links are connected with probability $p = 0.4$. Every new coming node makes $m=4$ connections, where $75\%$ connections are intra community and $25\%$ are inter community. Each community has almost equal number of nodes in the final generated network. In experiments, we have taken around $1\%$ nodes of innermost shells as core nodes. So, value of $q$ is $0.01$. Probability $r$ decides the density of the network, we have set $r=0.006$.

\begin{figure}[]
\centering
\includegraphics[width=8cm]{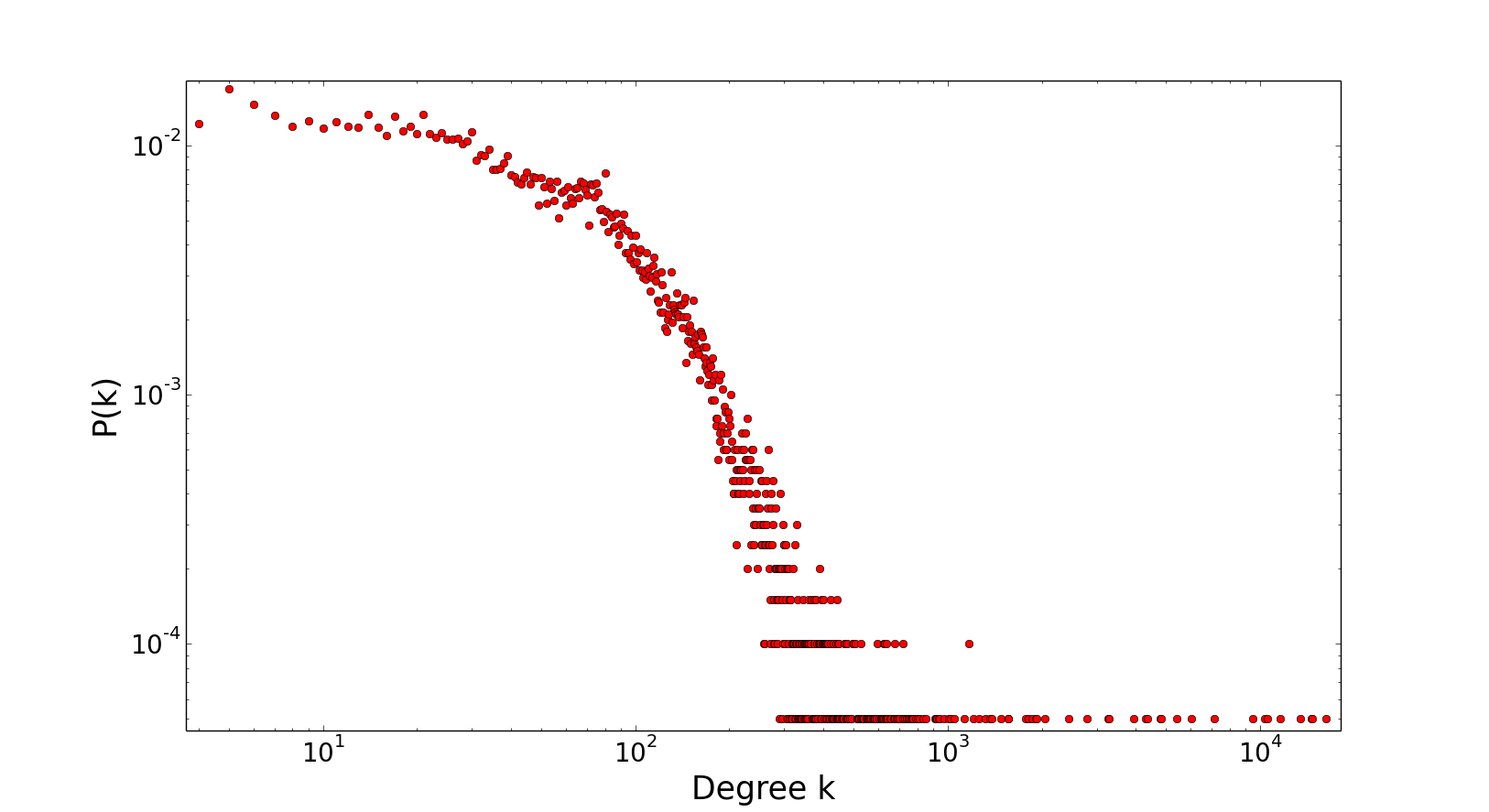}
\caption{Degree distribution of unweighted network}
\end{figure}

\begin{figure}[]
\centering
\includegraphics[width=8cm]{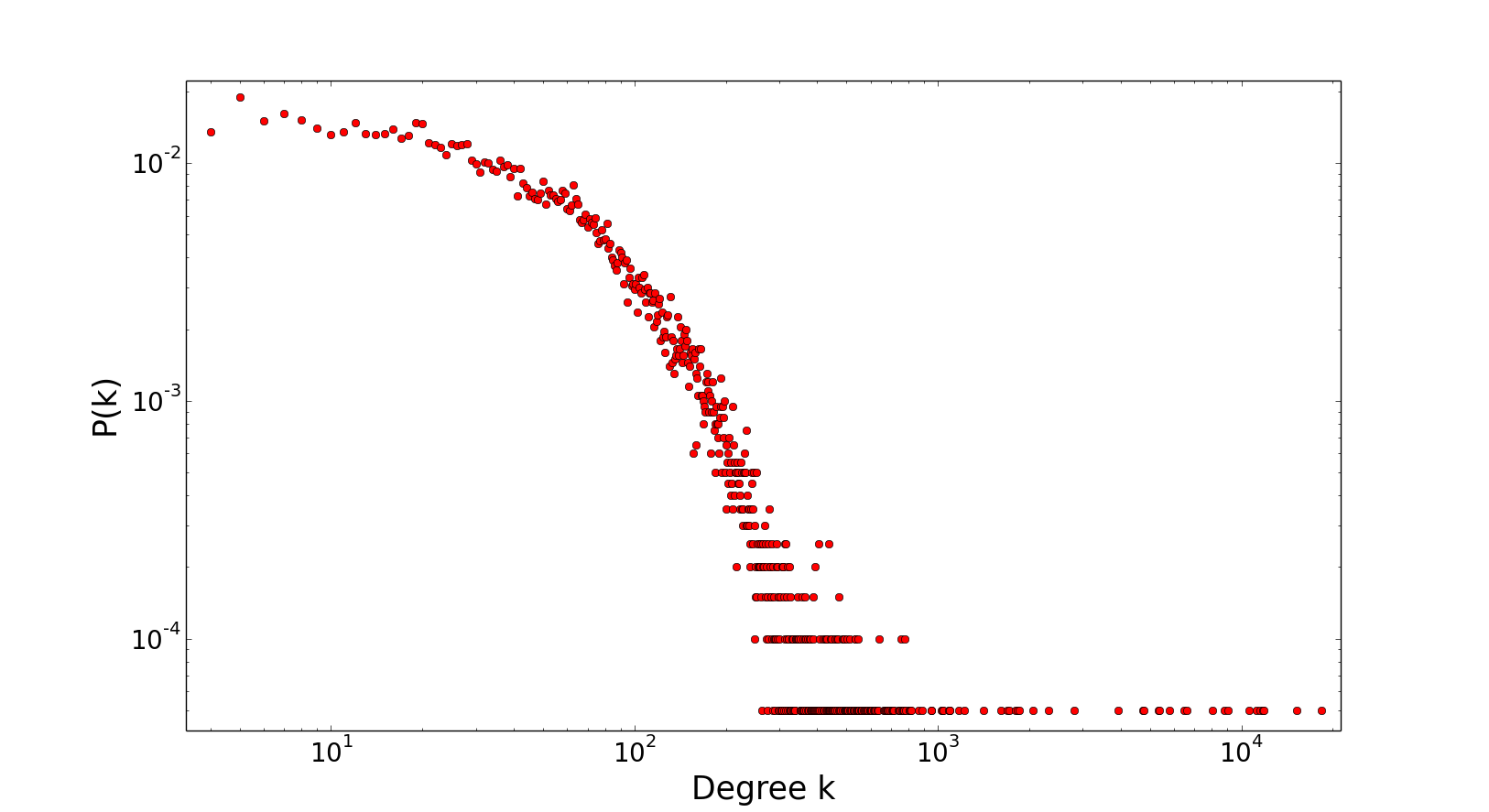}
\caption{Degree distribution of weighted network}
\end{figure}

In Fig. 5, we show that the degree distribution of the generated network follows power law, as $P(k) \propto k^{-\gamma}$ with exponent $\gamma \approx 2.5$. This has a droop head and a heavy tail. We further study average clustering coefficient and average nearest neighbor degree in the generated network. Clustering coefficient of a node shows how tightly knit a node is in its neighborhood. It is defined as,
\begin{center}
$CC(x) = \frac{2}{(k_x) * (k_x - 1)}\sum a_{xy}a_{yz}a_{xz}$
\end{center}
In Fig. 10, green color squares show average clustering coefficient with respect to degree of the nodes.

Average nearest neighbor degree represents assortativity of the networks. It can be calculated as,
\begin{center}
$k_{nn}(x) = \frac{1}{k_x}\sum_{y \epsilon \Gamma(x)} k_y$
\end{center}
In Fig. 11, green color squares denote average nearest neighbor degree $(k_{nn})$ for the generated network. This plot shows, as degree increases average nearest neighbor degree decreases. All results are very similar to real world networks.

\subsection{Model B}

We propose an evolving model for weighted networks by extending model A. Network starts with a seed graph similar to model A, having $c$ number of communities, where each community has a clique of size $n_0$. Each edge is assigned weight $w_0=1$. All nodes of the seed graph are set as core nodes. At every time stamp, the steps followed are:

\begin{enumerate}
\item A new periphery node $x$ is added to a uniformly at randomly selected community. This new node is connected with $m \cdot f$ intra community nodes using intra-community strength preferential attachment, and $m \cdot (1-f)$ inter community nodes using inter-community strength preferential attachment.

When a new node $x$ is connected to node $y$, we use BBV weight rearrangement function to balance the weights. This function distributes the weight across all neighbors of $y$, $z \epsilon \Gamma(y)$, proportional to their edge weights. It is defined as,
\begin{center}
$w_{yz}=w_{yz}+ \delta\frac{w_{yz}}{s_y}$
\end{center}
where, $\delta$ is a constant and represents the extra added load because of a new connection.
\item With probability $q$, a periphery node is selected using strength preferential attachment rule and it is converted into a core node. This node is connected to each core node with probability $p$.
\item Each node makes triad links with probability $r$ using strength preferential attachment. If node is already connected with chosen node then edge weight is increased by $w_0$.
\end{enumerate}

All new edges are assigned initial weight $w_0=1$ and probability parameters are same as defined in model A. Different values of $\delta$ can be chosen based on the evolving environment. In traffic networks, it denotes extra traffic load added to an existing node. In friendship networks, it shows, as time passes, a person makes new friends and also strengthens his relationships with old friends. Similarly, in collaboration networks, it represents that when a researcher collaborates with new people, he will also keep publishing with existing collaborators. $\delta \approx 1$, when all new coming load is distributed across all neighbors. $\delta < 1$, when a new node does not affect much the other existing neighbors. It can happen in friendship or collaboration kind of networks. $\delta > 1$, when a new edge bursts out more traffic to other connected nodes.

\subsection{Simulation Results}

\begin{figure}[]
\centering
\includegraphics[width=8cm]{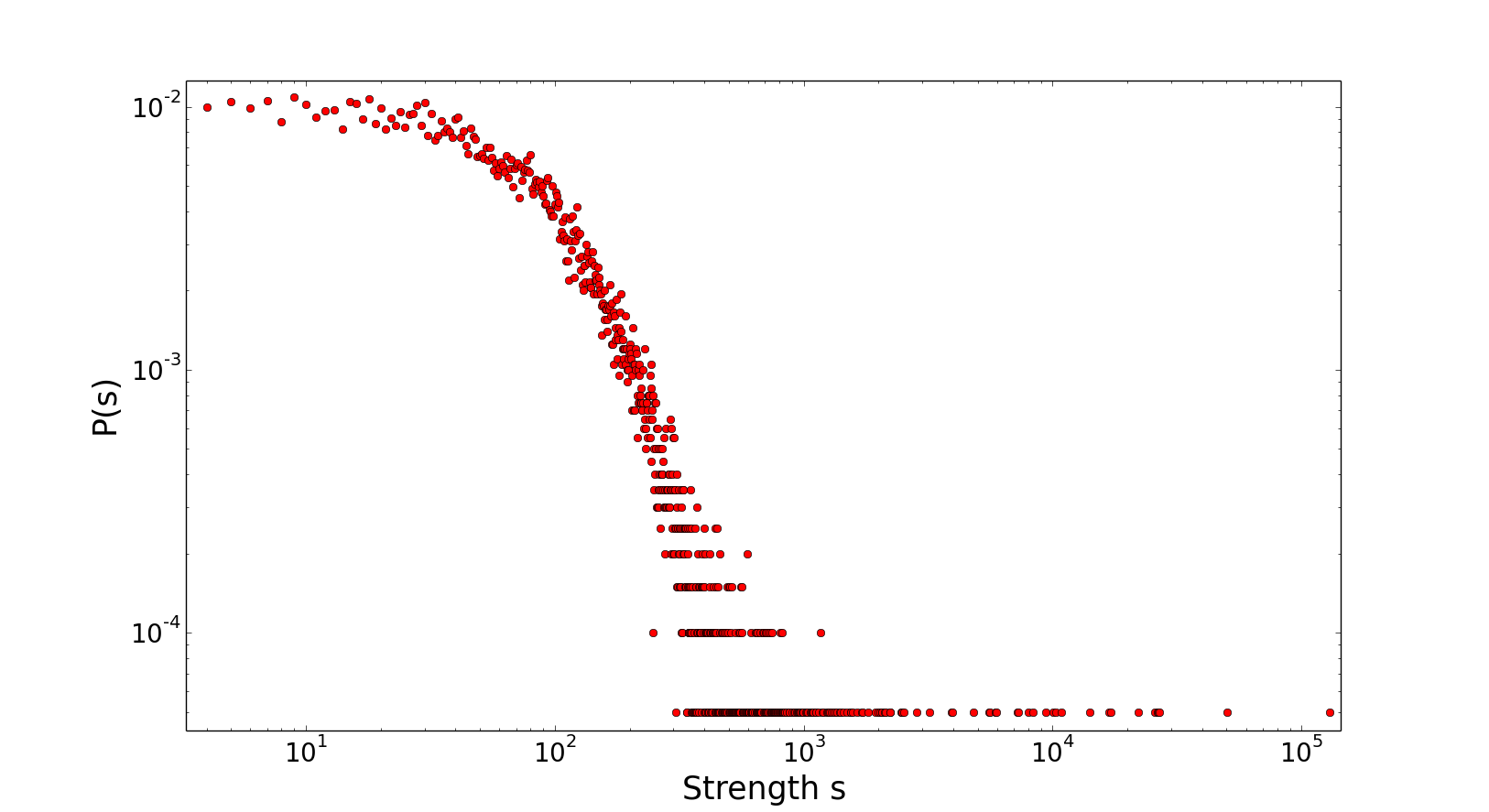}
\caption{Strength distribution}
\end{figure}

\begin{figure}[]
\centering
\includegraphics[width=8cm]{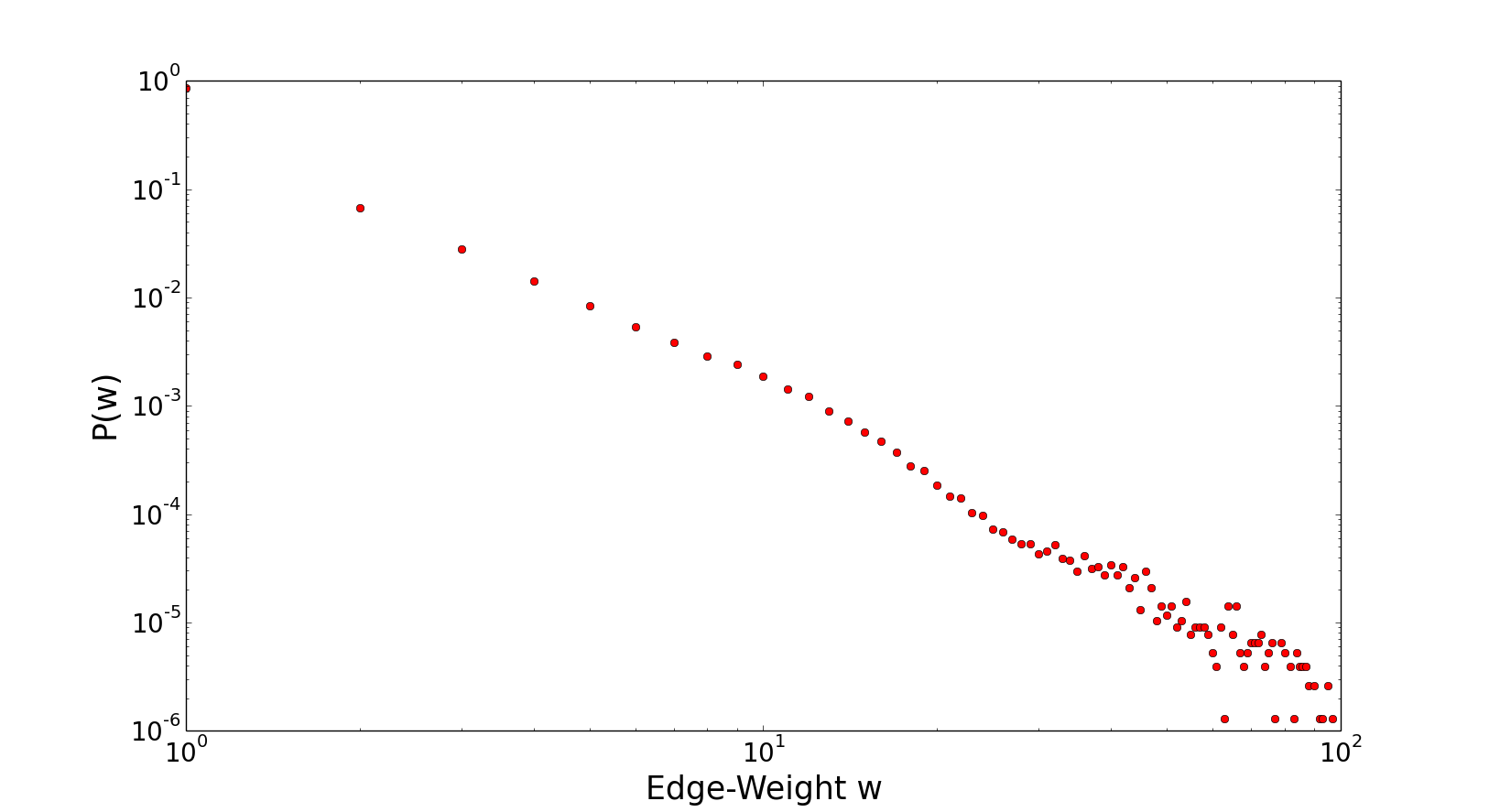}
\caption{Edge-weight distribution}
\end{figure}

We study properties of the generated networks by simulating the proposed model. Model is validated for different values of all the parameters. Results are shown for the network having 20,000 nodes and 768,875 edges. Values of the other parameters are set as, $c=6$, $n_0=3$, $m=4$, $f=0.75$, $p=0.4$, $q=0.01$, $r=0.006$, and $\delta=1$.

All results are plotted on log-log scale. In Fig. 6, Degree distribution P(k) of the generated network is plotted. It follows power law $P(k) \propto k^{-\gamma}$, where $\gamma \approx 2.2$. In Fig. 7, we plot the node strength distribution, that is defined as, the probability that a randomly selected node has strength value $s$. This distribution has a fat heavy tail and follows power law $P(s) \propto s^{-\gamma}$ with best fitted $\gamma \approx 2.3$. The edge-weight distribution also follows power law with exponent $\gamma \approx 3.5$ as shown in Fig. 8. It shows that strong or busy links keep getting stronger and busier. All plots show similar behavior as obtained for real world weighted networks. Fig. 9 shows that there is high correlation between average strength and degree of the nodes.
For weighted networks, weighted clustering coefficient can be calculated as,
\begin{center}
$C^w(x) = \frac{1}{(k_x) * (k_x - 1)}\sum (w_{xy}w_{xz}w_{yz})^{1/3}$
\end{center}
In Fig 10., blue color triangles show the average clustering coefficient and red color circles show average weighted clustering coefficient for the network.
Similarly, average weighted nearest neighbor degree can be defined as,
\begin{center}
$k^w_{nn}(x) = \frac{1}{s_x}\sum_{y \epsilon \Gamma(x)} w_{xy}k_y$
\end{center}
In Fig 11., average nearest neighbor degree and average weighted nearest neighbor degree is represented by blue triangles and red circles respectively.

\begin{figure}[]
\centering
\includegraphics[width=8cm]{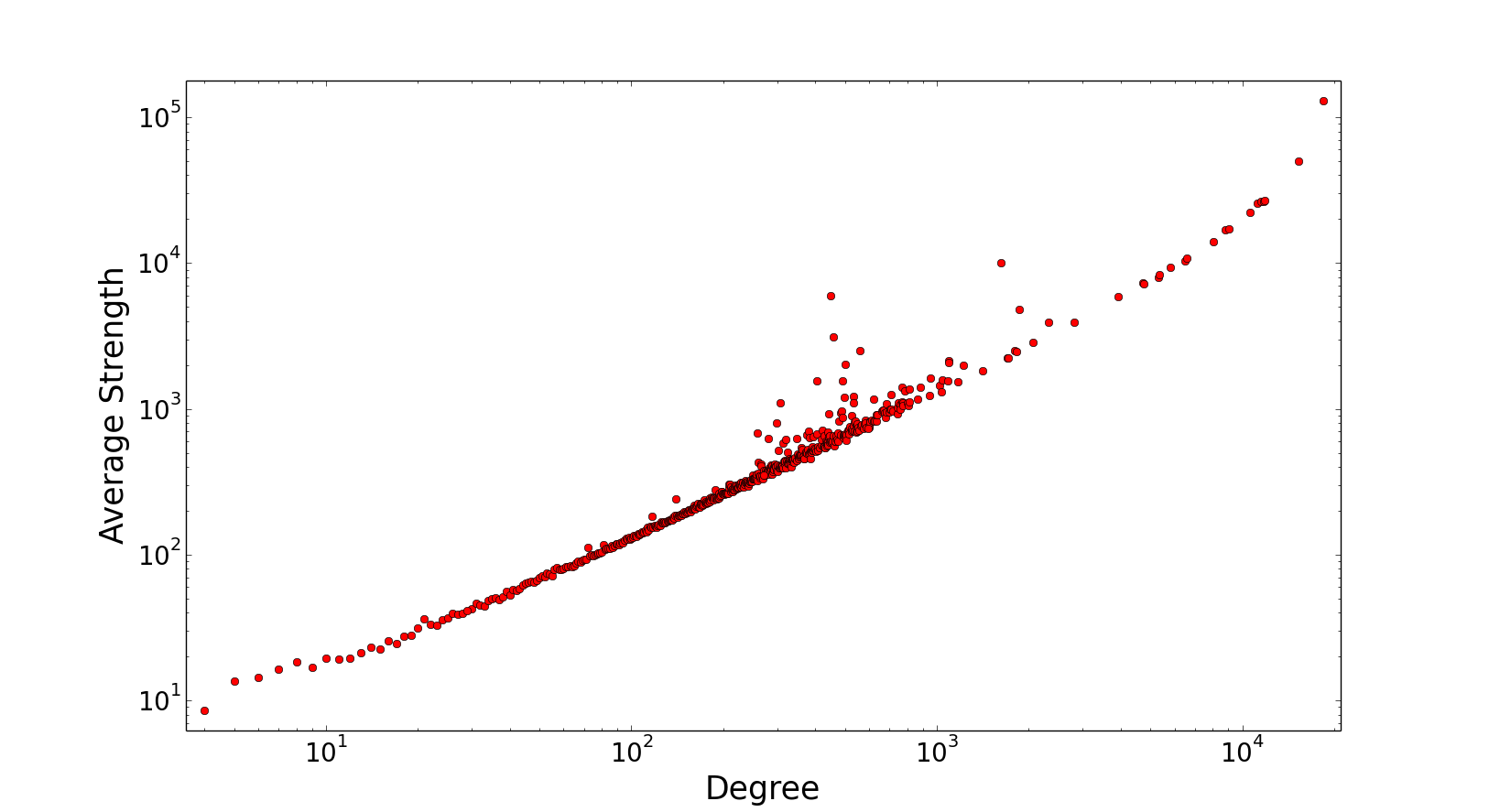}
\caption{Strength-Degree Correlation}
\end{figure}

\begin{figure}[]
\centering
\includegraphics[width=8cm]{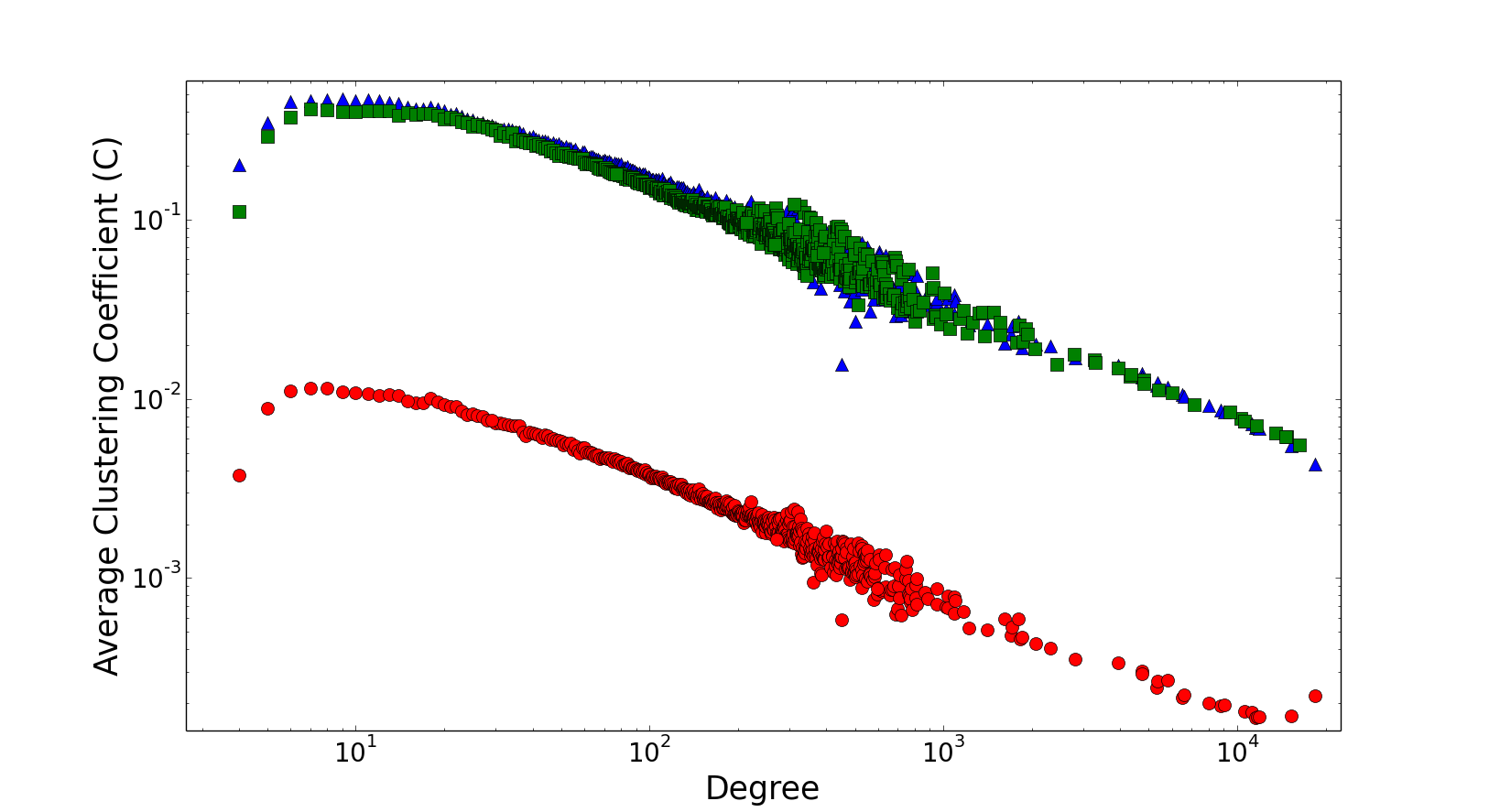}
\caption{Average Clustering Coefficient}
\end{figure}

\begin{figure}[]
\centering
\includegraphics[width=8cm]{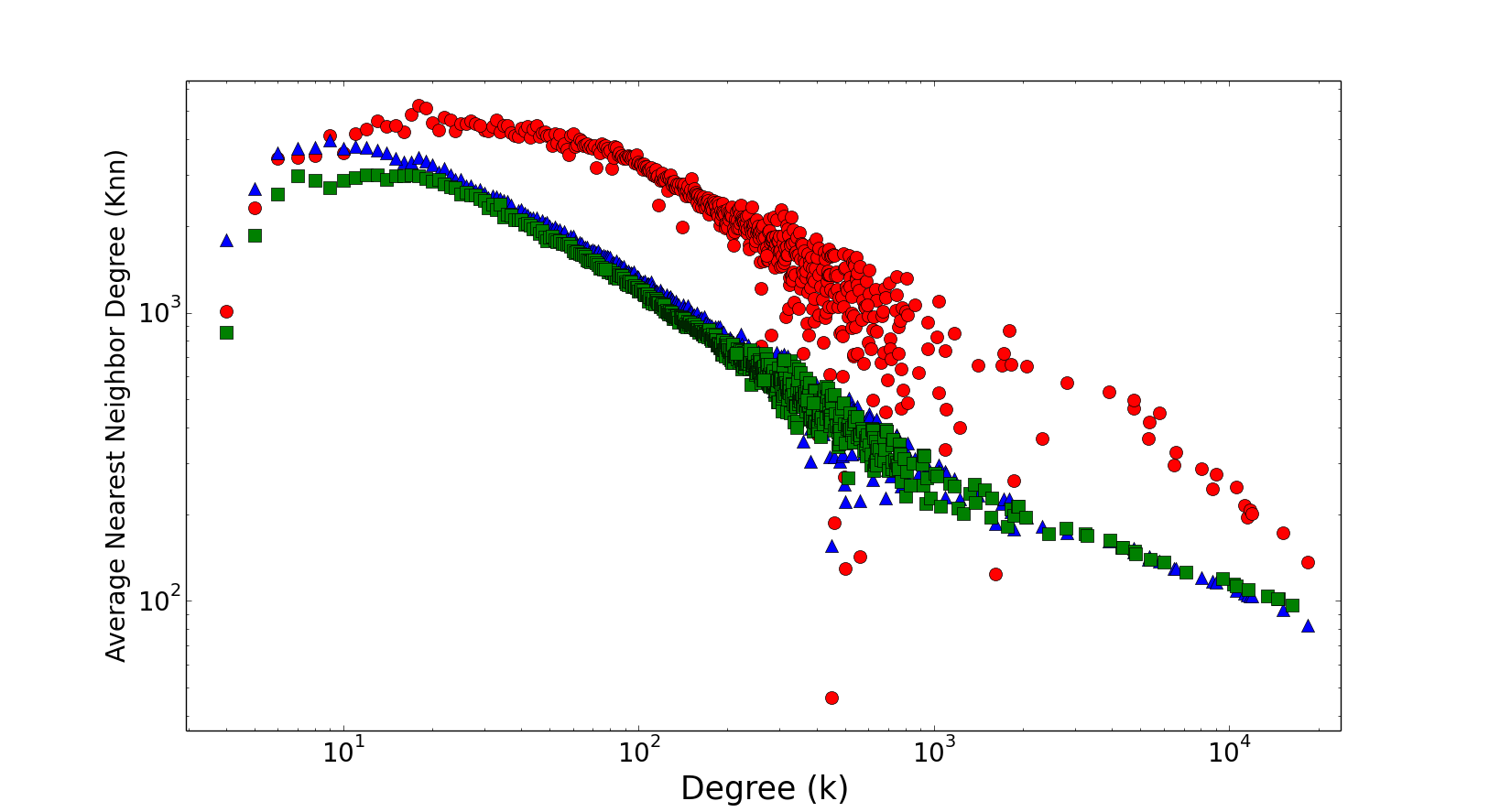}
\caption{k-Nearest Neighbor Degree}
\end{figure}

\subsection{Validation of the Proposed Models}

In the simulation, approximately $1\%$ innermost nodes are considered as core nodes. To validate the efficiency of our models, we detect core nodes in the generated networks. We use K-shell and S-shell decomposition methods for unweighted and weighted networks respectively. We calculate what percentage of nodes are misidentified. It includes the nodes which are marked as Core but do not belong to core or which are not marked as Core but belong to core. It can be defined as, 
\begin{center}
$Error = \frac{Number\ of\ misidentified\ nodes}{Total\ core\ nodes} * 100 \%$
\end{center}
So, efficiency of the models can be calculated as $(100 - Error) \%$. Table 2 shows the efficiency of the proposed models for different size of unweighted and weighted networks. Each value is calculated by averaging over 5 generated networks. In simulation, all probabilities are taken same as explained in simulation results sections except $\delta = 0.6$. Similar results are observed for different set of probabilities. In both models, efficiency increases with network size. In weighted networks, efficiency decreases as we increase value of $\delta$. This happens because for high value of $\delta$, nearest neighbors of core nodes receive more distributed strength and become part of inner shells close to the nucleus. 

\begin{table}[]
\centering
\caption{Efficiency of the proposed models}
\begin{adjustbox}{width=.5\textwidth}
\label{my-label1}
\begin{tabular}{|l|l|l|l|l|}
\hline
Number of nodes & \multicolumn{1}{|p{3cm}|}{\centering for Unweighted Networks\\(in \%)} & \multicolumn{1}{|p{3cm}|}{\centering for Weighted Networks\\(in \%)} \\ \hline
10000 & 96.94 & 96.33 \\ \hline
20000 & 98.02 & 96.49 \\ \hline
30000 & 98.35 & 97.13 \\ \hline
40000 & 98.54 & 97.56 \\ \hline
50000 & 98.83 & 98.02 \\ \hline
\end{tabular}
\end{adjustbox}
\end{table}


\section{Conclusion}

In the present paper, we have studied evolution process of the core in real world networks. We show that when a node is converted into a core node, increase in the number of connections with core nodes is more important than increase in the total degree. It helps a node to get a well known position in the network. We have proposed evolving models for networks having core-periphery and community structure. The growth dynamics of a network is based on topological growth and self growth. In weighted model, we use weight distribution function to balance the increased load. K-shell and S-shell decomposition methods are used to validate the efficiency of proposed models. It is shown that models generate core-periphery structure and marks core nodes with very less error.

Simulation results show that the generated networks follow all properties of real world networks. Unweighted networks follow power law degree distribution with a droop head and a flat heavy tail. Similarly, weighted networks follow degree, strength and edge-weight power law distributions such as statistically observed in real world weighted networks. In weighted networks, degree and strength is highly correlated. Furthermore, we analyze other properties of these networks, such as clustering coefficient and average nearest neighbor degree. Model parameters can be varied to get different clustering coefficient.

This work is a good contribution towards the field of modeling. We will do further analytical analysis of evolving equations for different network properties. These networks can be used to study dynamic phenomena taking place on real world networks like information diffusion, epidemic spread, spread of influence etc. In future, we will study these processes on the generated synthetic networks.


\bibliographystyle{IEEEtran}
\bibliography{mybib}

\end{document}